\documentclass[          
  reprint,
  aps,
  prd,
  superscriptaddress,
   nofootinbib
]{revtex4-2}

\usepackage[table]{xcolor}
\definecolor{todoblue}{RGB}{0, 91, 187}
\definecolor{accent1}{HTML}{DB9D47}
\definecolor{accent2}{HTML}{8CC084}
\definecolor{todoblue}{RGB}{0, 91, 187}

\usepackage{graphicx}
\usepackage{tikz}
\usepackage{subcaption}
\usepackage{multirow}
\usepackage{parskip}
\usetikzlibrary{shapes.geometric, arrows.meta, positioning}

\newcommand{\mbfcvar}[1]{%
    \textbf{\{#1\}}%
}

\usepackage{bm}

\newcommand{\mbfcvarsecond}[1]{%
  \ensuremath{\bm{\langle}}%
    \textbf{#1}%
  \ensuremath{\bm{\rangle}}%
}

\usepackage{xurl}  
\usepackage{hyperref}
\hypersetup{
  hypertexnames=false,
  colorlinks=true,
  allcolors=todoblue,
  urlcolor=todoblue,  
  linkcolor=todoblue,
  citecolor=todoblue,
  pdfborder={0 0 0},
  breaklinks=true,
}

\begin{abstract}

    Media Bias/Fact Check (MBFC) purports to quantify the bias, credibility, and factuality of reporting for roughly 10,000 media sources, and the resulting data is commonly used in misinformation research.
    In the present study, we show that MBFC's methodology does not meet basic standards of rigor for academic research.
    Despite its widespread prevalence, studies using MBFC rarely examine it carefully, often describing it in ways that contradict its ``About'' page, or treating it as authoritative despite MBFC's disclaimer that it is ``not a tested scientific method... [but] a simple guide to the idea of a source's bias.'' 
    We identified no papers that adequately describe MBFC as the opinions of a single person or critically engage with its methodology in order to justify proceeding with its use. 
    We argue that MBFC's data is not neutral or accurate, but a computationally legible account of hegemony, a specious dataset for uncritical research that mistakes the familiarity of the concepts it quantifies with accuracy.
    Our study concludes with a call for academic researchers to stop using MBFC. 
    MBFC's data quantifies the results of political processes, including campaigns to discredit the press, and presents them as simple facts about the world, thus reproducing the crisis misinformation scholarship exists to address.

\end{abstract}

\begin{document}

\title{
Stop using Media Bias/Fact Check in research
}

\author{
  \firstname{Alejandro Javier}
  \surname{Ruiz Iglesias}
}
\email{alejandro.ruiz@uvm.edu}

\affiliation{
  Vermont Complex Systems Institute,
  University of Vermont,
  Burlington,
  VT 05405,
  USA.
}
\affiliation{
  Computational Story Lab,
  University of Vermont,
  Burlington,
  VT 05405,
  USA.
}

\author{
  \firstname{Julia Witte}
  \surname{Zimmerman}
}
\affiliation{
  Vermont Complex Systems Institute,
  University of Vermont,
  Burlington,
  VT 05405,
  USA.
}
\affiliation{
  Computational Story Lab,
  University of Vermont,
  Burlington,
  VT 05405,
  USA.
}

\author{
  \firstname{Ashley~M.~A.}
  \surname{Fehr}
}
\affiliation{
  Vermont Complex Systems Institute,
  University of Vermont,
  Burlington,
  VT 05405,
  USA.
}
\affiliation{
  Computational Story Lab,
  University of Vermont,
  Burlington,
  VT 05405,
  USA.
}

\author{
  \firstname{Peter Sheridan}
  \surname{Dodds}
}
\affiliation{
  Vermont Complex Systems Institute,
  University of Vermont,
  Burlington,
  VT 05405,
  USA.
}
\affiliation{
  Computational Story Lab,
  University of Vermont,
  Burlington,
  VT 05405,
  USA.
}
\affiliation{
  Santa Fe Institute,
  1399 Hyde Park Rd,
  Santa Fe,
  NM 87501,
  USA
}

\affiliation{
  Complexity Science Hub,
  Metternichgasse 8,
  1030 Vienna,
  Austria
}

\author{
  \firstname{Christopher~M.}
  \surname{Danforth}
}
\affiliation{
  Vermont Complex Systems Institute,
  University of Vermont,
  Burlington,
  VT 05405,
  USA.
}
\affiliation{
  Computational Story Lab,
  University of Vermont,
  Burlington,
  VT 05405,
  USA.
}

\maketitle
\section{Introduction}
\label{introduction}

Social media has ushered in the so-called post-truth era, and, in response, the burgeoning field of misinformation science. As in all scientific endeavors, as part of its inquiry, the field must count, classify, and measure the properties of misinformation, the infrastructure for which is expanding. We examine a popular piece of that infrastructure, Media Bias/Fact Check (MBFC). 

MBFC is widely accepted in scientific practice~\cite{quote2, quote3,quote4,cinelli2021echo,infodemic, quote, justwebsite, wunschBigDataDemography2024}, recommended as a tool by university libraries~\cite{odhner2024mediabias, wichita2020evaluatingnews, uscupstate2025newsliteracy}, bundled into composite datasets~\cite{gallottiAssessingRisksInfodemics2020,gruppiNELAGT2020LargeMultiLabelled2021,lin2023correspondence,norregaardNELAGT2018LargeMultiLabelled2019}, and used in commercial products (such as the popular consumer news aggregation site ground.news~\cite{groundnews}). \citet{wunschBigDataDemography2024} aptly summarize this consensus in a footnote, writing ``MBFC is currently the most comprehensive media bias resource on the internet,'' an almost identical statement to that on MBFC's own homepage~\cite{mbfc}.

MBFC rates, among other things, the \mbfcvar{bias}\footnote{
    To avoid confusion between, for example, the general concept of bias and the MBFC variable by that same name, variables from MBFC will be written like \mbfcvar{this}, and their potential values like \mbfcvarsecond{this}.
} of media institutions along a left-to-right political spectrum. We examine MBFC's methodology and data (Section \ref{sec:data}) and show that they are unsuitable for academic use. We then create a database of papers, quantifying the presence of MBFC in the misinformation literature (Section~\ref{sec:quantifying}), and explore why authors use MBFC (Section~\ref{sec:why}), paying special attention to the many published descriptions of MBFC that contradict its own website (Section~\ref{sec:misconceptions}).

We argue that MBFC's data is not accurate, but specious. MBFC's data quantifies hegemony, the common sense ideas popularized by a ruling class to maintain their status~\cite{germanideology,gramsci1971prison}. More specifically, \mbfcvar{bias} is rooted in a political project to neutralize an adversarial media (Section~\ref{sec:history}). Both MBFC and the scholarship that relies on it are misled by political forces guiding their inquiry without their conscious awareness~\cite{Kolakowski78, germanideology}, causing researchers to conflate the familiarity of MBFC's conclusions with accuracy (Section~\ref{sec:stop}). This shared ideology allows MBFC to bypass examination, as demonstrated by the widespread mischaracterization of MBFC throughout the literature.

We conclude with a call to stop using MBFC in research (Section~\ref{sec:conclusion}). That MBFC's data is prevalent throughout the misinformation literature is a testament both to the success of the attacks on the press discussed in Section~\ref{sec:history} and to the failure of much misinformation scholarship to account for this political context.

\section{MBFC's Methodology and Data}
\label{sec:data}

MBFC has a ``Methodology'' page~\cite{Methodology2026} that outlines the various rubrics used to generate its data. The page has changed considerably since it was initially published. Until 2025, however, the page was thin~\cite{MBFCMethodology2019}. To our reading, it functions more like a ``Discussion'' section, with explanations about interpreting the scores. The 2024 version of the page~\cite{MBFCMethodology2024} contains the following example:

\begin{quote}
    Biased Wording = 4 (CNN uses moderately biased words that favor the left and headlines typically match the story)
    
    Factual/Sourcing = 4 (CNN has failed fact checks and sometimes omits critical information from stories to favor their perspective.)
    
    Story Choices/Editorial = 9 (CNN almost always favors pro-liberal stories and publishes negative conservative stories)
    
    Political Affiliation = 5 (CNN’s ownership owns other left-leaning outlets and favors Democratic Candidates)

Total = 22\\
Average 22/4 = 5.50\\
5.50 = Moderate Left Bias\\
\end{quote}

It then links to a page titled ``Left vs. Right Bias: How we rate the bias of media sources,'' which more closely resembles a typical ``Methodology'' section, whose substance consists of 804 words. It explains that...

\begin{quote}

    ...a source rated either right or left favors almost all of the policies in their category, whereas a Left-Center or Right-Center source will favor more of one side, but not all. Least Biased sources tend to be more balanced, provide perspectives from both sides, and have limited editorial positions.
  \end{quote}

  It then goes through the following ``policies,'' explaining the left and right position on those in a couple sentences: ``General philosophy,'' ``Abortion'', ``Economic policy'', ``Environmental policy'', ``Gay rights'', ``Gun rights'', ``Health care [sic]'', ``Immigration'', ``Military'', ``Personal responsibility'', ``Regulation'', ``Social views '', ``Taxes'', and ``Worker's/Business Rights.''

There is no further explanation. Why or how MBFC chose these specific topics goes without discussion.

In 2024, we start to see the beginnings of the rubrics that we will discuss at length in this section, e.g., references to \mbfcvar{economic system}~\cite{MBFCMethodology2024}. In 2025, MBFC redesigned the rating systems, making it much more highly specified with rubrics and scoring. This basic structure is the one in use in 2026~\cite{MBFCMethodology2025, Methodology2026}. 

\begin{table*}[!ht]
\centering
\begin{subtable}[t]{0.48\linewidth}
    \centering
    \begin{tabular}{|p{1cm}|p{1.5cm}|p{5.5cm}|}
    \hline
    \textbf{Score} & \textbf{Rating} & \textbf{Description} \\
    \hline
    0 & Very High & Consistently factual, uses credible information, no failed fact checks. \\
    \hline
    0.1--1.9 & High & High factual, minor sourcing issues, reasonable fact check record. \\
    \hline
    2.0--4.4 & Mostly Factual & Generally reliable but may have occasional fact-check failures, transparency, and sourcing issues. \\
    \hline
    4.5--6.4 & Mixed & Reliability varies; multiple fact-check failures, poor sourcing, lack of transparency, one-sidedness. \\
    \hline
    6.5--8.4 & Low & Often unreliable; frequent fact-check failures and significant issues with sourcing, transparency, propaganda, conspiracies, and pseudoscience promotion. \\
    \hline
    8.5--10 & Very Low & Consistently unreliable, heavily biased, with intentional misinformation likely. \\
    \hline
    \end{tabular}
\end{subtable}
\hfill
\begin{subtable}[t]{0.48\linewidth}
    \centering
    \begin{tabular}{|p{1cm}|p{7.5cm}|}
    \hline
    \textbf{Score} & \textbf{Description} \\
    \hline
    0 & Perfect balance, presenting all sides equally with no discernible bias or use of emotional language. \\
    \hline
    1 & Almost perfectly balanced, with very minor favoritism and minimal use of subtle emotional cues, but all perspectives fairly represented. \\
    \hline
    2 & Minor bias, slightly favoring one side with occasional use of emotionally suggestive terms, while still maintaining reasonable balance and representation of opposing views. \\
    \hline
    8 & Heavy bias, consistently favoring one side with little effort to present alternative viewpoints and pervasive use of emotional language that borders on propaganda. \\
    \hline
    9 & Strong bias, rarely including alternative perspectives, using extreme emotional framing or language that seeks to manipulate reader perception almost entirely in favor of one viewpoint. \\
    \hline
    10 & Extreme bias or propaganda, exclusively presenting one side with no balance or acknowledgment of opposing views, often employing inflammatory, divisive, or manipulative language to an extreme degree. \\
    \hline
    \end{tabular}
\end{subtable}
\caption{MBFC's guides for interpreting \mbfcvar{factual reporting} (left) and \mbfcvar{one-sidedness} (right). For the latter, we remove rows 3--7 for brevity.}
    \label{tab:factual-reporting}
\end{table*}

When discussing MBFC's methodology, we discuss only the most up to date version, as of this writing. However, it is important to note that most of the papers found in Section~\ref{sec:quantifying} predate the present rubric, when MBFC's lack of methodological rigor was impossible to dispute. 

For the rest of this section, we discuss MBFC's \mbfcvar{factual reporting}, \mbfcvar{bias}, \mbfcvar{credibility}, and \mbfcvar{freedom}. The first three are available through a single API endpoint, which returns a CSV file rating $9,365$ sources. MBFC provides these scores for a diverse set of outlet's, but the typical example is a media organization, such as CNN or Fox News, though one also finds unions and think tanks. The \mbfcvar{freedom} score is for countries, not sources, and it comes directly from the MBFC site and is not available through the site's API.
For important context, we discuss each of these concepts' design and score paradigms. 
\begin{table*}[t]
\centering
\begin{subtable}[t]{0.48\linewidth}
    \centering
    \begin{tabular}{|p{1.8cm}|p{4.1cm}|c|}
    \hline
    \textbf{Category} & \textbf{Rating} & \textbf{Points} \\
    \hline
    \multirow{5}{1.8cm}{Factual Reporting}
    & Very High & 4 \\
    & High & 3 \\
    & Mostly Factual & 2 \\
    & Mixed & 1 \\
    & Low & 0 \\
    \hline
    \multirow{4}{1.8cm}{Bias}
    & Least Biased / Pro-Science & 3 \\
    & Right-Center or Left-Center & 2 \\
    & Left or Right & 1 \\
    & Questionable/Conspiracy/\newline Pseudoscience & 0 \\
    \hline
    \multirow{4}{1.8cm}{Traffic/ Longevity}
    & High Traffic & 2 \\
    & Medium Traffic & 1 \\
    & Minimal Traffic & 0 \\
    & Bonus: $\geq 10$ years existence & +1 \\
    \hline
    \multirow{2}{1.8cm}{Press Freedom}
    & Limited Freedom & -1 \\
    & Total Oppression & -2 \\
    \hline
    \end{tabular}
    \label{tab:credibility_rubric}
\end{subtable}
\hfill
\begin{subtable}[t]{0.48\linewidth}
    \centering
    \begin{tabular}{|p{3cm}|p{4.35cm}|}
    \hline
    \textbf{Credibility Level} & \textbf{Criteria} \\
    \hline
    High Credibility & A score of 6 or above. \\
    \hline
    Medium Credibility &
    A score between 3--5 points. Additionally, in accordance with the MBFC scoring system, a ``Mostly Factual'' rating with a score between 3.6 and 4.5 automatically results in a Medium Credibility classification, regardless of the overall tally in other categories. This reflects the critical importance of factual accuracy in determining credibility. \\
    \hline
    Low Credibility &
    A score of 0--2 points. Sources rated as ``Questionable,'' ``Conspiracy,'' or ``Pseudoscience'' are automatically classified as Low Credibility. \\
    \hline
    \end{tabular}
    \label{tab:cred-class}
\end{subtable}
\caption{MBFC's scoring rubric (left) and interpretation criteria (right) for \mbfcvar{credibility}. Note that, in the top left, \mbfcvar{factual reporting}'s lowest score (\mbfcvarsecond{very low}) is also omitted in the original. Note also the nested logic on the right,  which overrides the parent rubric.}
\label{tab:credibility}
\end{table*}

\subsection{Factual reporting}

According to MBFC's methodology page, the \mbfcvar{factual reporting} score is based on a 10-point scale, created using the following rubric: ``Failed Fact Checks (40\%), Sourcing (25\%), Transparency (25\%), and One-Sidedness/Omission (10\%).'' In what is a theme throughout this section, there is no way to access the original scale. The API only provides access to the ratings presented in Table~\ref{tab:factual-reporting}, from \mbfcvarsecond{very low} to \mbfcvarsecond{very high}, and the transformation is irregularly binned. The bin size for \mbfcvarsecond{high}, for example, is from 1--1.9, but \mbfcvarsecond{mostly factual} is from 2.0--4.4. No explanation is given for this transformation.

Table~\ref{tab:factual-reporting} shows the \mbfcvar{one-sidedness} portion of the \mbfcvar{factual reporting} score. Despite MBFC reporting an ostensibly separate \mbfcvar{bias} score (which we discuss in Section~\ref{sec:bias}), 10\% of the \mbfcvar{factual reporting} score contains what is conceptually difficult to distinguish from \mbfcvar{bias}. We discuss this in Section~\ref{sec:discussion}. 

Finally, a small note on notation: When plotting \mbfcvar{factuality}, we use a 0--5 scale, where 0 is \mbfcvarsecond{very low} and 5 is \mbfcvarsecond{very high}. This is an inverted scale from MBFC's internal one, in which smaller numbers correspond to higher factuality, but, since we do not have access to those numbers anyway, we have chosen the more intuitive alternative.

\subsection{Credibility}
\label{credibility}

The methodology for calculating \mbfcvar{credibility} is shown in Table~\ref{tab:credibility}. It is a composite of \mbfcvar{factual reporting}, \mbfcvar{bias}, and \mbfcvar{freedom}, along with the site's traffic, for which they use an estimate of page views. Though this process generates a numerical score, we are given access to a binned version with uneven bin sizes described in Table~\ref{tab:credibility}. Note that this table contains nested logic, with 2 of the branches resulting in an override to what is described in the parent rubric.

\subsection{Freedom}
\label{Freedom Index}

MBFC assigns each country a \mbfcvar{freedom} score. The methodology is as follows~\cite{mbfc_country_methodology_2023}:

\begin{quote}
    We calculate the overall freedom levels of countries by averaging the scores from Reporters Without Borders (RSF) and Freedom House’s yearly Freedom in the World Index [...] If a country has not been rated by either source, we use other sources such as Unesco.org, Statista.org, BBC Country Profiles, or other credible sources to estimate the level of freedom.
\end{quote}

The methodology uses a point score system from 0--100. However, when downloading the data, the original 100-point score is unavailable. Instead, it is grouped into five categorical labels: 
Total oppression (0--24),
Limited freedom (25--49),
Moderate freedom (50--69),
Mostly free (70--89), and
Excellent freedom (90--100).

The uneven bucket size is unexplained on the methodology page~\cite{mbfc_country_methodology_2023}, and the raw score is unavailable through the API. Thus, available data is on a five-point scale of uneven bin size in the original rating space. For our analysis, we label these bins 0--4, with 0 being ``Total Oppression'' and 4 being ``Excellent Freedom.''

We consider the freedom scores for ``countries with which the United States has strained or hostile relations'', including: Afghanistan, Belarus, China, Cuba, Eritrea, Iran, Myanmar, Nicaragua, North Korea, Russia, South Sudan, Syria, and Venezuela~\cite{EnemiesUnitedStates2026}. MBFC rates every one of these countries' freedom as ``total oppression'' except for one, South Sudan, which reaches ``limited freedom''. Similarly, NATO countries have a mean score of $3.1$ whereas non-NATO have a mean score of $2.0$.

\subsection{Bias}
\label{sec:bias}

\begin{table}[tb]
\begin{subtable}{\linewidth}
    \begin{tabular}{|c|p{5.6cm}|}
    \hline
    Score Range & Bias Category \\
    \hline
    $-10$ to $-8.0$   & Extreme Left Bias \\
    $-7.9$ to $-5.0$  & Left Bias (Far Left at $-7.0+$ [sic]) \\
    $-4.9$ to $-2.0$  & Left-Center Bias \\
    $-1.9$ to $+1.9$  & Least Biased \\
    $+2.0$ to $+4.9$  & Right-Center Bias \\
    $+5.0$ to $+7.9$  & Right Bias (Far Right at $+7.0+$) \\
    $+8.0$ to $+10$   & Extreme Right Bias \\
    \hline
    \end{tabular}
    \caption{MBFC rates its \mbfcvar{bias} scores from $-10$ to $+10$; communicates the score in its data in a five-point scale, but explains their methodology on this seven-point scale. The categories \mbfcvarsecond{left bias} and \mbfcvarsecond{right bias} contain internal subdivisions, presumably corresponding to the breaks in the five-point scale used in the API data. Note that ``Far Left at $-7.0+$'' is a direct quote. The ``$+$'' is, we believe, meant to indicate ``more to the left''.}
    \label{tab:bias-schema}
\end{subtable}

\vspace{0.8em}

\begin{subtable}{\linewidth}
    \centering
    \begin{tabular}{|c|p{7cm}|}
    \hline
    Score & Description \\
    \hline
    $-10$ & \textbf{Communism}: Advocates no corporatism, extreme regulation, and full government ownership of industries. \\ \hline
    $-7.5$ & \textbf{Socialism}: Supports minimal corporatism, high regulation, and significant government ownership. \\ \hline
    $-5$ & \textbf{Democratic Socialism}: Endorses reduced corporatism with strongly regulated capitalism. \\ \hline
    $-2.5$ & \textbf{Regulated Market Economy}: Promotes moderate corporatism with balanced regulations. \\ \hline
    $0$ & \textbf{Centrism}: Balances regulation and corporate influence without significant bias. \\ \hline
    $2.5$ & \textbf{Moderately Regulated Capitalism}: Leans slightly toward corporatism with moderate government intervention. \\ \hline
    $5$ & \textbf{Classical Liberalism}: Emphasizes moderate to high corporatism with lower regulations. \\ \hline
    $7.5$ & \textbf{Libertarianism}: Advocates low government intervention and high corporate influence. \\ \hline
    $10$ & \textbf{Radical Laissez-Faire Capitalism}: Advocates minimal to no regulation, with the economy governed entirely by free-market principles and private enterprise. \\ \hline
    \end{tabular}
    \caption{MBFC calculates \mbfcvar{bias} with a rubric, with 35\% of that score coming from \mbfcvar{economic system}. MBFC ``[e]valuates the economic ideology the source promotes, from communism to no-regulation capitalism''~\cite{mbfc}.}
    \label{tab:economic-system}
\end{subtable}
\caption{MBFC's \mbfcvar{bias} categories and score definitions.}
\end{table}

\begin{figure}[t]
    \centering
    \includegraphics[width=\linewidth]{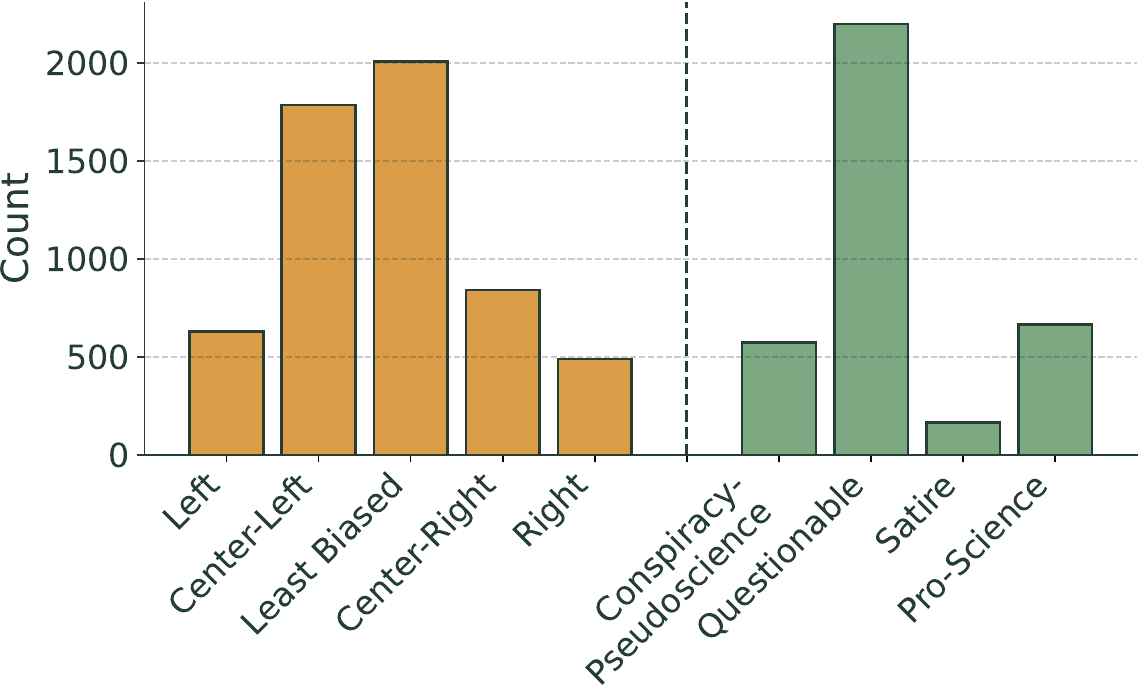}
\caption{
 Distribution of \mbfcvar{bias} categories in MBFC. Categorical values outside the left/right spectrum are separated out to simplify interpretation. Note that the two are exclusive, e.g., a source cannot be both \mbfcvarsecond{least biased} and \mbfcvarsecond{pro-science}, or \mbfcvarsecond{center-right} and \mbfcvarsecond{questionable}.
    }
\label{fig:bias}

\end{figure}

Figure~\ref{fig:bias} shows counts of sources in MBFC's dataset grouped by their \mbfcvar{bias} ratings. This scale presents problems of interpretation. The left-to-right values are, as one might expect, ordinal values that can be arranged along a one-dimensional spectrum, but it is difficult to explain or understand why \mbfcvarsecond{pro-science}, \mbfcvarsecond{questionable}, \mbfcvarsecond{satire}, and \mbfcvarsecond{conspiracy-pseudoscience}, categorical values outside that spectrum, appear as potential \mbfcvar{bias} values.

\begin{figure}[t]
    \centering
    \includegraphics[width=\linewidth]{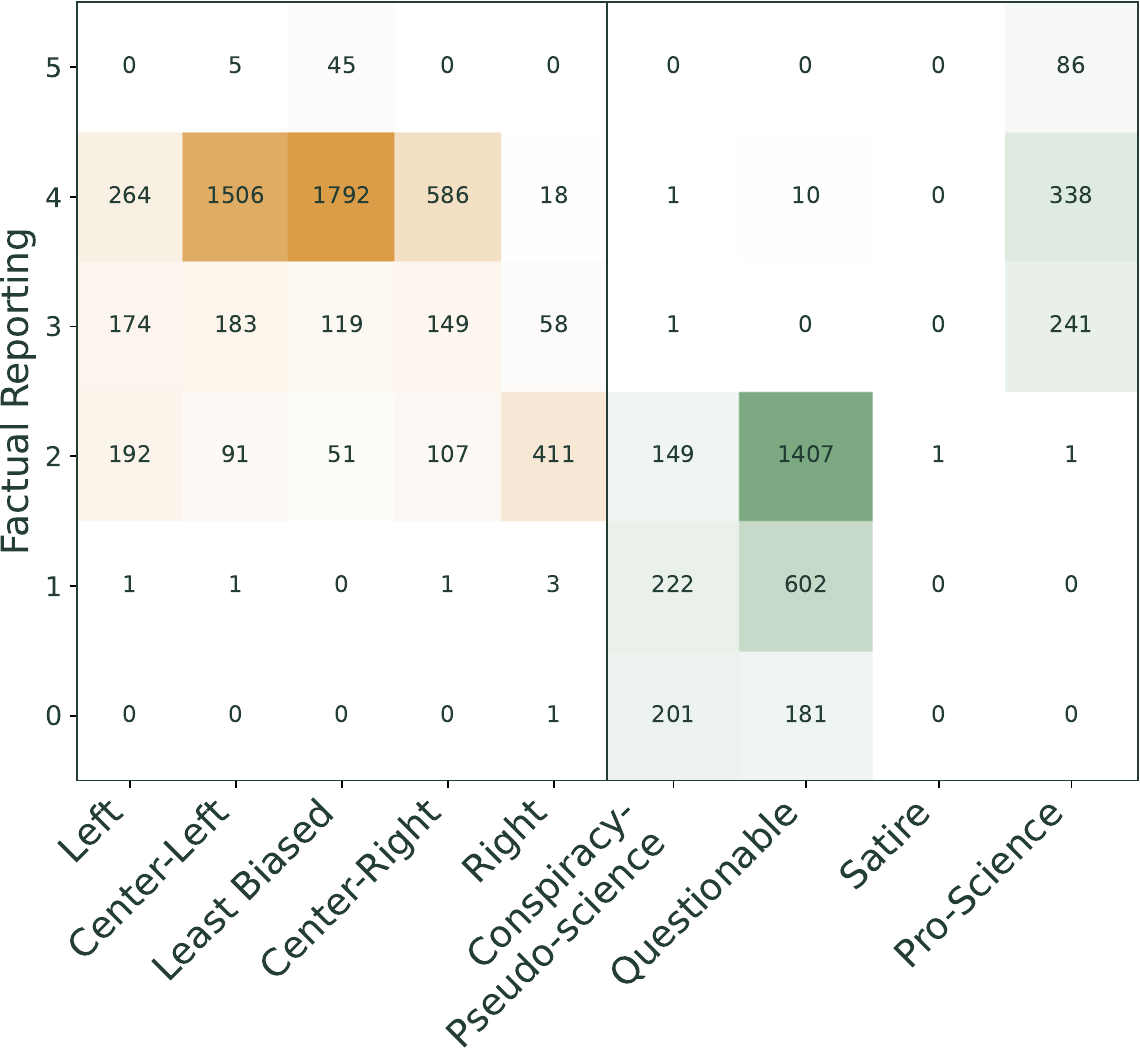}
\caption{
 \mbfcvar{bias} vs. \mbfcvar{factual reporting}. The categorical values outside the left/right spectrum are to the right, represented in green, whereas the left/right values are to the left in gold.
 }
\label{fig:bias_v_fact}
\end{figure}

Figure~\ref{fig:bias_v_fact} also compares \mbfcvar{bias} to \mbfcvar{factual reporting}. Taken together with Figure~\ref{fig:bias}, researchers might find a familiar idea, that of the ``liberal media'', which we discuss further in Sections~\ref{sec:history} and \ref{sec:conclusion}.

Further complicating the interpretation, MBFC's methodology page (see Table~\ref{tab:bias-schema}) does not include these extra categories. It shows instead a \mbfcvar{bias} score that goes from $-10$ to $+10$, but which is then binned into a 7-point scale. The actual data exposed to the user, however, contains a 5-point scale (once the values mentioned above are removed). The original $-10$ to $+10$ scale seems internal to MBFC and is never exposed to users. Within the explanation of the 7-point scale, however, the second and second-to-last items contain parentheticals which we believe are intended to explain the 5-point scale. 

To summarize, there are three different scales used: a twenty-point scale, a seven-point scale, and a five-point scale. The twenty-point scale seems to be the original, internal scale, which gets coarse grained into seven- and five-point scales. The seven-point scale seems to exist only in the explanation for the bias score. The five-point scale is potentially explained in parentheticals in that same methodology section, and it is the data with which MBFC's users directly interact, with the caveat that it also contains the four extra values that do not sit anywhere on the spectrum.

For some clarity on these four extra values, we can turn to the \mbfcvar{credibility} rubric (discussed in Section~\ref{credibility} and in Table~\ref{tab:credibility}). Recall that \mbfcvar{credibility} is a composite score that includes \mbfcvar{bias}. A source is more credible if it is either \mbfcvarsecond{least biased} or \mbfcvarsecond{pro-science}, for which it earns 3 points. If it is \mbfcvarsecond{right-center} or \mbfcvarsecond{left-center}, it gets 2 points. For \mbfcvarsecond{left} or \mbfcvarsecond{right}, 1 point; and for \mbfcvarsecond{questionable/conspiracy/pseudoscience}, which we assume is a composite of \mbfcvarsecond{questionable} and \mbfcvarsecond{conspiracy-pseudoscience}, it gets 0 points.

To calculate \mbfcvar{bias}, MBFC uses the following scoring categories~\cite{Methodology2026}:
\begin{quote}
    The placement of a source on the Left-Right Bias Scale is determined by a weighted composite score derived from four categories: Economic System (35\%), Social Progressive Liberalism vs. Traditional Social Conservatism (35\%), Straight News Reporting Balance (15\%), and Editorial Bias (15\%). Scores are on a scale of $-10$ to $+10$, and the weighted average determines the overall bias score.
\end{quote}

We provide two scoring category examples from the full \mbfcvar{bias} rubric. First, the \mbfcvar{economic system} rubric portion, accounting for the plurality of the score at 35\%, can be found in Table~\ref{tab:economic-system}. In it, MBFC labels the farthest left as ``Communism,'' defined as ``[advocating] no corporatism, extreme regulation, and full government ownership of industries.'' The farthest right is ``Radical laissez-faire capitalism,'' which ``[a]dvocates minimal to no regulation, with the economy governed entirely by free-market principles and private enterprise.'' 

Second, the \mbfcvar{straight news reporting balance} rubric portion...

\begin{quote}
    [m]easures how well a source reports all sides in its straight news stories, either through story selection or content balance within articles. This covers strictly news reporting and is separate from Editorial/Op-Ed bias.
\end{quote}

We discuss these examples further in Section~\ref{sec:discussion}.

\section{MBFC in the literature}
\label{sec:quantifying}

A systematic analysis of MBFC in the scientific literature is difficult because not all papers that rely on MBFC cite or credit it in the same way. To illustrate this issue, searching citations databases Semantic Scholar~\cite{ammar2018literaturegraph}, OpenAlex~\cite{priem2022openalex}, OpenCitations~\cite{peroni2019opencitations}, and Scopus~\cite{scopus} 
for ``mediabiasfactcheck.com'' or ``Media Bias/Fact Check'' yields relatively few results. For these strings, respectively, 
Semantic Scholar shows 2 and 21 results, 
OpenCitations shows 0 and 0, 
OpenAlex shows 7 and 0, and 
Scopus shows 1 and 142 results. 
Meanwhile, Google Scholar returns $1,610$ results for ``Media Bias/Fact Check.'' 

The discrepancy between Google Scholar and the various metadata searches suggests that many papers that use MBFC as a dataset do not have its citation in their references, at least not in the format the metadata-based citation databases are expecting.

We choose not to use the results from Google Scholar for two reasons. First, its indexing and algorithm are opaque, so much so that we find the total number of results for the same query fluctuates day to day. Second, we find its permissiveness towards web scraping similarly variable. We conclude that these factors would harm the reproducibility of our work.

More importantly, citations represent the official scientific account of how knowledge spreads. One of our main arguments is that MBFC's methodology is flawed, and that, through its continued use in research, MBFC's scores become divorced from its methodology such that the provenance is effectively erased. Therefore, we use citations because they are the mechanism by which readers of studies ought to be able to trace such processes. 

\subsection{Methods}
\label{sec:quantifying.methods}

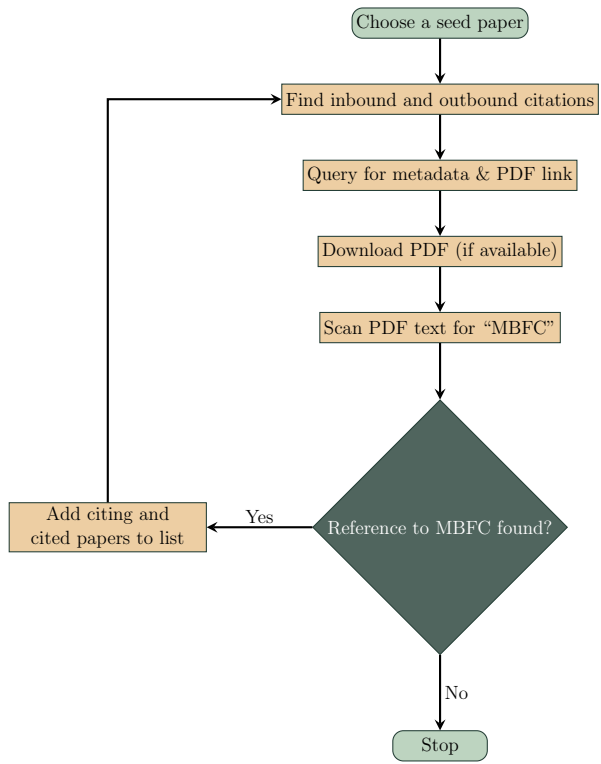
\begin{figure}[h]
\definecolor{green1}{HTML}{7CA982}
\definecolor{orange1}{HTML}{DB9D47}
\definecolor{dark1}{HTML}{243E36} 
\definecolor{white1}{HTML}{FFFFFF} 
\tikzset{     
startstop/.style={rectangle, rounded corners, minimum width=2.5cm, minimum height=0.8cm, text centered, draw=dark1, fill=green1!50, font=\Large},     
process/.style={rectangle, minimum width=3cm, minimum height=0.8cm, text centered, draw=dark1, fill=orange1!50, font=\Large},     
decision/.style={diamond, minimum width=2.5cm, minimum height=0.8cm, text centered, draw=dark1, fill=dark1!80, font=\Large, text=white},     
arrow/.style={thick,->,>=stealth} 
} 
\begin{tikzpicture}[
    node distance=1.2cm and 2.5cm, scale=0.5, transform shape
    ]      
\node (start) [startstop] {Choose a seed paper};     
\node (step1) [process, below=of start] {Find inbound and outbound citations};     
\node (step2) [process, below=of step1] {Query for metadata \& PDF link};     
\node (step3) [process, below=of step2] {Download PDF (if available)};     
\node (step4) [process, below=of step3] {Scan PDF text for ``MBFC''};     
\node (decide) [decision, below=of step4, yshift=-0.3cm] {Reference to MBFC found?};  
\node (loop) [process, text width=5cm, left=of decide, xshift=-.3cm] {Add citing and cited papers to list};     
\node (stop) [startstop, below=of decide, yshift=-0.8cm] {Stop};      
\draw [arrow] (start) -- (step1);     
\draw [arrow] (step1) -- (step2);     
\draw [arrow] (step2) -- (step3);     
\draw [arrow] (step3) -- (step4);     
\draw [arrow] (step4) -- (decide);     
\draw [arrow] (decide.south) -| (stop.north) node[near end,right,font=\Large]{No};     
\draw [arrow] (decide.west) -- (loop.east) node[midway,above,font=\Large]{Yes};     
\draw [arrow] (loop.north) -- ++(0,2) |- (step1.west);     
\end{tikzpicture}  
\caption{The snowball sampling process looks at the text of a paper. If it finds a reference to MBFC, it fetches all its references and recursively begins the process anew on each one.}
\label{snowballdiagram}
\end{figure}

To estimate the prevalence of MBFC in the scientific literature, we use snowball sampling (see Figure~\ref{snowballdiagram}). We seed this snowball sampling process with the papers exported from the Scopus search for ``Media Bias/Fact Check.'' For each seed paper, we create a Paper object in a database. We then fetch every paper that our seed paper cites (``outbound citations'') or that cites our seed paper (``inbound citations''). For the fetched papers, we also create Paper objects, which we link to the seed papers through a Citation object, a directional link between the two papers. 

We then attempt to fetch the manuscript text of each fetched paper. If successful, we scan this text for mentions of MBFC. If the paper contains MBFC, we add it to our seed papers and recursively repeat the algorithm, again outlined below for clarity:

\begin{enumerate}
    \setlength{\itemsep}{0pt}
    \setlength{\parskip}{0pt}
    \item   For each paper, find both inbound and outbound citations.
    \item   For both the resulting inbound and outbound citations, use metadata   databases OpenAlex, Crossref, and OpenCitations to find metadata and a link to a PDF version of the paper.
    \item   Fetch the PDF, if possible.
    \item   Scan the manuscript text of the paper for references to MBFC.
    \item   If the paper contains a reference to MBFC, return to step 1.
\end{enumerate} 

We then run the sampler to its natural end, i.e., until no new papers are found to re-seed the algorithm.

\subsection{Results}
\label{sec:quantifying.results}
 
We find MBFC present in 3.50\% ($372$) of the $10,642$ papers in our sample (or 5.0\% of papers if we exclude those for which we could not fetch the manuscript text; see Table~\ref{tab:misinfo-papers}). 
We expect that 3.50\% is an underestimate of the impact of MBFC in the misinformation literature for the following reasons.

\begin{table*}[th]
\centering
\begin{tabular}{|l|p{.2\textwidth}|p{.2\textwidth}|p{.2\textwidth}|p{.2\textwidth}|}
\hline
& \multicolumn{2}{|c|}{\textbf{Entire database}}
& \multicolumn{2}{|c|}{\textbf{Papers containing ``misinformation''}} \\
\cline{2-5}
& Papers with text & All papers
& Papers with text & All papers \\
\hline
Papers with MBFC & 329 (5.01\%) & 372 (3.50\%)
                  & 259 (7.38\%) & 263 (6.49\%) \\
Total papers     & 6,514 & 10,642
                  & 3,509 & 4,051 \\
\hline
\end{tabular}
\caption{Counts of papers with MBFC in the entire database and in the subset of papers containing the term ``misinformation'' in their title, abstract, or body.}
\label{tab:misinfo-papers}

\end{table*}
\textbf{No paper text available}: First, for 61\% of papers, we are unable to fetch the actual manuscript text, terminating the snowball sample for that branch (even though subsequent connected papers could have been relevant to this analysis). Similarly, detecting MBFC in a paper involves scanning its text, meaning that there are almost certainly papers in our database that use MBFC that we cannot observe. Even in the case that we do find the PDF, scanning its text comes with many limitations, including problems with formatting and inconsistent OCR.

\begin{figure}[t]
    \begin{center}
    \includegraphics[width=\linewidth]{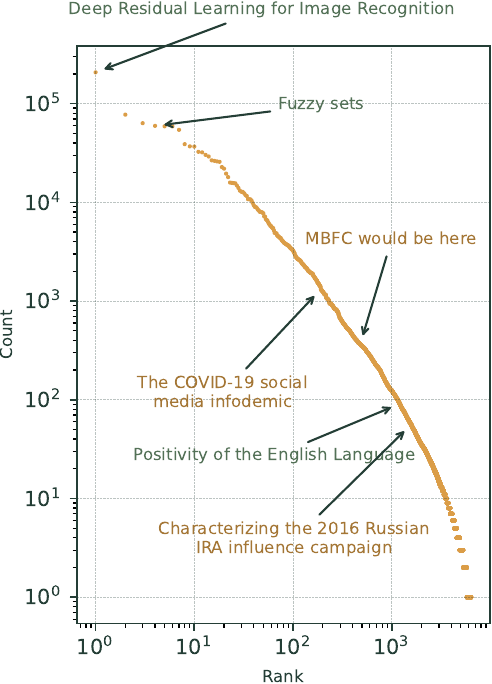}
    \caption{Rank by citation count, with the most cited paper as rank $=1$. Exponential decay is fast ($\alpha=1.93$), reflecting our snowball algorithm's preference for highly-cited papers. Titles of papers whose manuscript text contains MBFC are in \textcolor{accent1!70!black}{gold}. Those without are in \textcolor{accent2!60!black}{green}.}
    \label{fig:zipf}
    \end{center}
\end{figure}

\textbf{Greediness of snowball algorithm}: Further diluting the percentage, the greediness of the methodology biases our database towards highly cited papers. Figure~\ref{fig:zipf} shows the rank distribution of papers in our database. Most of the highly cited papers in our database are outside of the field of misinformation studies, diluting the percentage scores. Conversely, we suspect that there are papers in the literature with low citation counts using MBFC that we did not find.

\textbf{Bundled sources including MBFC}: Our study only takes into account studies that use MBFC directly. We know of at least 4 sources that combine MBFC with other data and some interpretative work to create composite databases~\cite{gallottiAssessingRisksInfodemics2020,gruppiNELAGT2020LargeMultiLabelled2021,lin2023correspondence,norregaardNELAGT2018LargeMultiLabelled2019}. As in the case of MBFC, these data sources can be difficult to trace through citations and metadata. A more complete accounting of MBFC's usage in the literature would require repeating the snowball process on a complete list of secondary sources of MBFC.

\textbf{Google Scholar}: We do not use Google Scholar's counts directly for the reasons outlined previously. By comparison to our count of $372$, Google Scholar returns roughly $4.3\times$ this count at $1,610$ results for the search query ``Media Bias/Fact Check.''

\begin{table}[!htbp]
  \centering
\footnotesize
\rowcolors{2}{accent2!2}{accent2!4}
\begin{tabular}{@{}
  p{.15\linewidth}@{}p{.15\linewidth}@{}p{.695\linewidth}@{}
}

\hline
Google Scholar Citations &
Open Citations Citations &
\multicolumn{1}{c}{\multirow{2}{*}{\parbox[c][1cm]{1cm}{Title}}}\\  
\hline
$3395$  &  $1320$  &  The Covid-19 Social Media Infodemic \\
$3824$  &  $845$  &  The Echo Chamber Effect On Social Media \\
$2092$  &  $630$  &  A Survey Of Fake News \\
$1487$  &  $570$  &  Influence Of Fake News In Twitter During The 2016 Us Presidential Election \\
$986$  &  $367$  &  Beyond News Contents \\
$775$  &  $361$  &  Assessing The Risks Of ‘Infodemics’ In Response To Covid-19 Epidemics \\
$787$  &  $249$  &  “Fake News” Is Not Simply False Information: A Concept Explication And Taxonomy Of Online Content \\
$976$  &  $241$  &  Auditing Radicalization Pathways On Youtube \\
$419$  &  $221$  &  Covid-19 Vaccine Hesitancy On Social Media: Building A Public Twitter Data Set Of Antivaccine Content, Vaccine Misinformation, And Conspiracies \\
$581$  &  $203$  &  Fighting An Infodemic: Covid-19 Fake News Dataset \\
$613$  &  $184$  &  Analyzing The Digital Traces Of Political Manipulation: The 2016 Russian Interference Twitter Campaign \\
$337$  &  $153$  &  The Stealth Media? Groups And Targets Behind Divisive Issue Campaigns On Facebook \\
$456$  &  $151$  &  FANG \\
$372$  &  $150$  &  Fake News Detection Based On News Content And Social Contexts: A Transformer-Based Approach \\
$325$  &  $127$  &  Sentiment Analysis For Fake News Detection \\
$284$  &  $125$  &  The Covid-19 Infodemic: Twitter Versus Facebook \\
$444$  &  $121$  &  Examining The Alternative Media Ecosystem Through The Production Of Alternative Narratives Of Mass Shooting Events On Twitter \\
$377$  &  $114$  &  You Are Fake News: Political Bias In Perceptions Of Fake News \\
$312$  &  $111$  &  Recovery \\
$215$  &  $99$  &  False News On Social Media \\
$207$  &  $83$  &  Evaluating Deep Learning Approaches For Covid19 Fake News Detection \\
$365$  &  $70$  &  Combating Disinformation In A Social Media Age \\
$143$  &  $67$  &  A Hybrid Linguistic And Knowledge-Based Analysis Approach For Fake News Detection On Social Media \\
$264$  &  $67$  &  Exploring The Role Of Visual Content In Fake News Detection \\
$167$  &  $64$  &  Search Bias Quantification: Investigating Political Bias In Social Media And Web Search \\
$148$  &  $64$  &  Who Falls For Online Political Manipulation? \\
$145$  &  $63$  &  Red Bots Do It Better:Comparative Analysis Of Social Bot Partisan Behavior \\
$200$  &  $60$  &  Nela-Gt-2018: A Large Multi-Labelled News Dataset For The Study Of Misinformation In News Articles \\
$153$  &  $59$  &  Overview Of Constraint 2021 Shared Tasks: Detecting English Covid-19 Fake News And Hindi Hostile Posts \\
$160$  &  $57$  &  Credibility-Based Fake News Detection \\
$186$  &  $56$  &  Characterizing The 2016 Russian Ira Influence Campaign \\
$101$  &  $50$  &  Understanding High- And Low-Quality Url Sharing On Covid-19 Twitter Streams \\
$96$  &  $50$  &  Health Misinformation Detection In The Social Web: An Overview And A Data Science Approach \\

\hline
\end{tabular}
\caption{Most cited papers using MBFC in our database with citations $\ge50$ according to \citet{peroni2019opencitations}. For additional context, we also include citation counts from Google Scholar~\cite{google_scholar}.}
\label{tablepapers}
\end{table}

In an attempt to better estimate MBFC's impact in the relevant literature, instead of looking at all papers uncovered by the sampling methodology described above, we include only those that additionally contain the word ``misinformation'' somewhere in the title, abstract, or manuscript text. In this subset, 4,051 papers of the original 10,642 remain, of which 263 (6.49\%) reference MBFC. If we further filter those to papers for which we have manuscript texts, we find 3,509, of which 259 (7.38\%) reference MBFC. Note that $372 - 259 = 113 $ papers contain MBFC but are filtered out of this subset.

In summary, in our database, we find MBFC present in $3.5\%$ to $7.4\%$ of the relevant literature, depending on how we filter. This is not marginal, but wisely used, and shows the importance of validating whether it is suitable for academic use (Section~\ref{sec:data}).

Table~\ref{tablepapers} contains the titles of papers in which we detected MBFC that have 50 or more citations, according Open Citations~\cite{peroni2019opencitations}. Of the 32 papers listed, 8 are developing methodologies for detecting misinformation, meaning that MBFC is used in the literature as a ground truth for machine learning models. If these models are subsequently used in other studies, or perhaps in industry, there is a risk of encoding MBFC's methodology and conclusions into future work. Due to aforementioned constraints in citation data, the original methodologies for those judgments, e.g., why a model might label something as ``fake news,'' are difficult to ascertain. We discuss this in Sections~\ref{sec:discussion} and \ref{sec:conclusion}.






\section{MBFC as described in the literature}
\label{sec:why}

In Section~\ref{sec:data}, we consider MBFC's data and methodology but, in Section~\ref{sec:quantifying}, we show that, despite its methodological flaws, authors still use MBFC widely. We address this tension here in two ways. First, we analyze how authors describe MBFC, arguing that they rarely examine MBFC closely. Second, we ask what authors seek to accomplish with MBFC's data: What is it that MBFC allows authors to do that makes it so appealing? 

\subsection{Methods}
\label{sec:why.methods}

We use a combination of examples from our reading and Natural Language Processing (NLP) methods. For the latter, we compile all papers that have MBFC mentions, then, for each paper, we combine the text from the title, abstract, and body of the paper. This text serves as the searchable document for MBFC mentions. Across sentences in these papers, we identify aliases of MBFC (e.g., ``MBFC'' or ``mediabiasfactcheck.com'').
When we find these aliases, we build a context window around each sentence containing an MBFC alias that includes 1 sentence before and 1 sentence after.

\begin{table}[b]
\centering
    \begin{tabular}{|p{.25\linewidth}|p{.7\linewidth}|}
    \hline
    Part of speech & Top words \\
    \hline
    Adjective & political, fake, factual, right, low, high, unreliable, extreme, top, reliable, social, left, questionable, independent, different. \\\hline
    
    Proper noun  & twitter, facebook, covid, \texttt{\_\_MBFC\_ALIAS\_\_}, march, wikipedia, april, alexa. \\\hline
    
    Noun & news, bias, media, source/sources, websites, outlets, domains, information, data, labels, articles, dataset, list, credibility. \\\hline
    
    Verb & used/using/use, left, based, shared, mixed, provided, labeled, obtained, listed, collected, published, classified, rated. \\\hline
    \end{tabular}
\caption{Part of speech top word counts for the context windows mentioning MBFC.}
\label{tab:alias-pos-counts}
\end{table}

We replace all aliases with the protected token \texttt{\_\_MBFC\_ALIAS\_\_}. We de-duplicate contexts, replace digits with \texttt{NUM}, and remove stopwords, including web and academic artifacts (e.g., ``http'', ``et'', ``al'', ``fig'', and roman numerals) to produce a bag of words count (see Table~\ref{tab:alias-analysis-bow}). Finally, we remove contexts containing a digit directly before an alias, as this most often indicates a footnote, which yield false contexts, since a footnote need not have any relationship to the footnote before or after. 

These cleaning steps remove $241$ contexts, leaving $728$ for analysis. Note that a single article may have multiple contexts. We tag parts of speech using Stanza~\cite{qi2020stanza}, then show the most frequent adjectives, proper nouns, nouns, and verbs occurring within our contexts in Table~\ref{tab:alias-pos-counts}.

\begin{table*}[t]
\centering
    \begin{tabular}{p{3.5cm}p{3.5cm}p{3.5cm}p{3.3cm}}
    \hline
    1-19 & 20-38 & 39-57 & 58-75 \\
    \hline
    news (801) & articles (116) & extreme (76) & facebook (57) \\
    \_\_MBFC\_ALIAS\_\_ (477) & dataset (116) & using (75) & ratings (57) \\
    bias (351) & used (116) & number (75) & shared (56) \\
    media (337) & based (115) & sites (74) & results (54) \\
    sources (306) & high (114) & domain (73) & social (54) \\
    political (247) & credibility (112) & twitter (72) & far (53) \\
    right (198) & website (110) & one (71) & urls (53) \\
    left (198) & tweets (103) & mixed (70) & scores (53) \\
    fake (181) & center (101) & top (69) & posts (52) \\
    websites (157) & leaning (97) & use (68) & outlet (51) \\
    factual (152) & users (96) & links (66) & biased (51) \\
    source (149) & also (95) & label (63) & questionable (50) \\
    outlets (148) & misinformation (95) & reliable (62) & independent (49) \\
    domains (146) & two (90) & categories (61) & provided (48) \\
    information (145) & content (89) & analysis (60) & online (48) \\
    data (144) & pages (89) & scale (60) & research (48) \\
    labels (134) & reporting (83) & lists (59) & accounts (47) \\
    low (122) & score (77) & category (58) & labeled (47) \\
    list (119) & unreliable (76) & fact (57) &  \\
    \hline
    \end{tabular}
\caption{Top words (count) in MBFC-containing articles within the context windows mentioning MBFC. The context windows include a single sentence prior to and after any MBFC alias.}
\label{tab:alias-analysis-bow}
\end{table*}

\subsection{Results}
\label{sec:why.results}

Consistent with our previous work~\cite{ruiz}, the proper nouns show that studies that mention MBFC focus on social media. In the adjectives, we see many of the descriptors of MBFC's rating systems discussed in Section~\ref{sec:data} (\mbfcvarsecond{low}, \mbfcvarsecond{high}, \mbfcvarsecond{left}, and \mbfcvarsecond{right} are \mbfcvar{credibility}/\mbfcvar{factual reporting} and \mbfcvar{bias} scores), and ``social'' describes its noun-pair ``media''. The rest of the nouns contain, similar to the proper nouns, various areas which authors use MBFC to study, like ``outlets'' and ``sources,'' as well as MBFC terms like \mbfcvar{bias}. In the verbs, there are examples of what MBFC provides the authors, e.g., sources are ``rated'' or ``classified.'' Combining these results, we see that studies turn to MBFC to study social media. They use it as a dataset to classify different kinds of sources by their credibility or bias.

From our reading and qualitative sampling of contexts, many texts describe MBFC as a ``fact-checking organization''~\cite{quote2, quote3, flaminoPoliticalPolarizationNews2023a,quote4}, often adding the word ``independent''~\cite{infodemic, quote}, or invoking it alongside scholarly sources, e.g.,~``various scholars and fact-checking organizations''~\cite{quote2}. In total, of the 372 papers that use MBFC, 42 contain the phrase ``fact-checking organization'', often describing MBFC directly. Sometimes, texts provide further details, describing MBFC as, for example, ``a small independent team of researchers and journalists''~\cite{flaminoPoliticalPolarizationNews2023a}. Others provide almost no details at all, and instead solely refer to it only by its URL~\cite{justwebsite}, without using the full name ``Media Bias/Fact Check''. 



~\citet{infodemic}, the most cited paper in our database, gives a typical example that is consistent with our NLP results and our reading:

\begin{quote}
    We tag links as reliable or questionable according to the data reported by the independent fact-checking organization Media Bias/Fact Check. In order to clarify the limits of an approach that is based on labeling news outlets rather than single articles, as for instance performed in [other studies], we report the definitions used in this paper for questionable and reliable information sources. In accordance with the criteria established by MBFC, by questionable information source we mean a news outlet systematically showing one or more of the following characteristics: extreme bias, consistent promotion of propaganda/conspiracies, poor or no sourcing to credible information, information not supported by evidence or unverifiable, a complete lack of transparency and/or fake news. By reliable information sources we mean news outlets that do not show any of the aforementioned characteristics. 
\end{quote}

This paragraph focuses mostly on what MBFC allows its authors to accomplish, methodologically speaking, but spends very little time discussing MBFC itself. As noted previously, \citet{infodemic} refer to MBFC as an ``independent fact-checking organization,'' as do many other texts in our database. We rarely get descriptions of MBFC itself beyond a simple sentence like this. Of the most used words in our context windows, only ``dataset,'' ``website,'' and ``independent'' are words that describe MBFC. 

The second most cited paper in our database has the same lead author. It describes MBFC with an identical quote and uses it similarly. 

In the third-most cited paper, ``A Survey of Fake News: Fundamental Theories, Detection Methods, and Opportunities,'' MBFC appears alongside similar datasets 
in a section titled ``Resources for Understanding News Publishers.''

\begin{quote}
    We introduce several resources that can help obtain the ground truth on the credibility (or political bias) of news publishers. One resource is the Media Bias/Fact Check website, which provides a list of media along with their political slant: left, left-center, least biased, right-center, and right.~\cite{surveyfakenews}
\end{quote}

This example recommends MBFC as a resource. Consistent with our NLP findings and previous examples, we again find no discussion of its methodology, only a description of its data.

As a final example, our fourth most cited paper adds some nuance to this discussion. \citet{justwebsite} provide descriptive statistics of their data (which combines MBFC and AllSides, a similar source):

\begin{quote}
    Using this final separation in seven classes, we identify in our dataset (we give the top hostname as an example in parenthesis): 16 hostnames corresponding to fake news websites (e.g. thegatewaypundit.com), 17 hostnames for extremely biased (right) news websites (e.g. breitbart.com), 7 hostnames for extremely biased (left) news websites (e.g. dailynewsbin.com), 18 hostnames for left news websites (e.g. huffingtonpost.com), 19 hostnames for left leaning news websites (e.g. nytimes.com), 13 hostnames for center news websites (e.g. cnn.com), 7 hostnames for right leaning websites (e.g. wsj.com), and 20 hostnames for right websites (e.g. foxnews.com).
\end{quote}

They also link to MBFC's methodology page in their description of MBFC.

The string ``mediabiasfactcheck.com/methodology'' appears 8 times in our database, which includes the references, making this a rare example of an author acknowledging that MBFC's data is the result of extensive interpretative work, the nature of which often goes unacknowledged in the literature.

\section{Discussion}
\label{sec:discussion}

\subsection{Misconceptions in the literature}
\label{sec:misconceptions}

As we discussed in Section~\ref{sec:quantifying}, though studies often use MBFC's data, they rarely describe MBFC. When there is a description, it is often inconsistent with how MBFC describes itself. Contradicting many of the examples in Section~\ref{sec:why}, MBFC's ``About'' page contains the following description:

\begin{quote} 
    Media Bias Fact Check, LLC is a North Carolina-based Limited Liability Company solely owned and operated by Dave Van Zandt. He makes all final editorial and publishing decisions.~\cite{mbfc}
\end{quote}  

Similarly contradicting previous descriptions, MBFC's author explicitly denies being a journalist: 

\begin{quote}
    Dave Van Zandt is a registered Non-Affiliated voter who values evidence-based reporting. Though not a journalist, Dave has maintained a lifelong interest in politics and media bias. He originally pursued a Communications degree in college before ultimately earning a degree in Physiology. Since then, he has worked in the healthcare industry (Occupational Rehabilitation) while continuing to study media, language, and bias independently.

    Over the past 20 years, he has studied media bias and linguistics and has applied the scientific method to create a structured, evidence-based methodology for assessing media bias and factual reporting.~\cite{mbfc}
\end{quote}  

Despite being used as, for example, ground truth in machine learning models~\cite{razaFakeNewsDetection2022}, or to definitively label sources as unreliable~\cite{infodemic}, MBFC provides the following disclaimer, which we interpret as being in tension with the previous quote:

\begin{quote}
    Disclaimer: The methodology used by Media Bias Fact Check is our own. It is not a tested scientific method. It is meant as a simple guide for people to get an idea of a source's bias.~\cite{mbfc}
\end{quote}

The descriptions in Section~\ref{sec:why} give the impression that MBFC is a professional ``fact-checking organization,'' conjuring the image of well-credentialed journalists at some kind of institution. MBFC is, as its website makes clear, one person's hobby. 
In our database, we have not been able to find a single paper that adequately describes MBFC as the opinions of one person or critically engages with its methodology in order to justify proceeding with its use despite its scientific limitations.

\subsection{History}
\label{sec:history}

Despite the framework's popularity, it is not obvious that the political bias of large institutions can be meaningfully distilled to a single point in one dimension. This particular conception of bias contains many complex, overlapping assumptions, both in what it includes and ignores. For example, MBFC's rubric focuses only on the content published by each institution, disregarding its governance and finances.  We therefore consider the historical and political context of its ideas here.

The roots of \mbfcvar{bias} can be traced to the Progressive Era reformers' arguments that the media had a systemic bias in favor of capital stemming from their corporate structure and/or the high cost of operating a media ~\cite{bauer2026making}. From there, this kind of structural critique was transformed, passing through McCarthyism and the civil rights movement, until becoming foundational to the coalescing conservative movement~\cite{greenbergIdeaLiberalMedia2008b}, which undertook a sustained public relations campaign to discredit the media for its supposed liberal bias~\cite{bauer2026making,Herman1999Myth}. 

From the 1950s through the 1960s, Southern segregationists, the John Birch Society, and Barry Goldwater supporters were coalescing into an anti-communist, proto-conservative movement, all of which had received critical coverage from an incipient national media~\cite{bauer2026making, greenbergIdeaLiberalMedia2008b}. Complimenting this bottom-up movement-building, there was a top-down-funded intellectual movement, beginning with oil tycoon H.L. Hunt's \textit{Facts Forum}, and including William F. Buckley's \textit{National Review}~\cite{bauer2026making}. From its inception, this rising (proto-)conservative media apparatus advocated that its readers learn to identify the bias of the mainstream press by analyzing the content of its stories~\cite{bauer2026making}. This nascent conservative coalition attacked the supposed liberal media, with some, like Archsegregationist George Wallace, using it to defend against its coverage of the civil rights movement, which was routinely downplayed in Southern papers~\cite{greenbergIdeaLiberalMedia2008b}. 

In 1969, with the ``liberal media'' critique establishing itself in the conservative movement, Reed Irvine founded Accuracy in Media (AIM), which ``helped make media bias an explicit conservative movement cause''~\cite{bauer2026making}. Originally, AIM spun off from the Council Against Communist Aggression (CACA), through which it attracted Morris Ernst, a prominent ACLU lawyer, and other liberal or progressive media reformers who brought with them the remnants of the Progressive Era's structural media criticism. Despite its CACA origins and initial putative bipartisanship, AIM's criticism of the media quickly found a constituency in the conservative movement. AIM became increasingly aligned with this movement as President Nixon and Vice President Agnew regularly attacked the media for its liberal bias~\cite{greenbergIdeaLiberalMedia2008b}, which, in their telling, interfered with the media's duty to tell the truth~\cite{bauer2026making}. Ernst, disillusioned, eventually left the organization, but only after the conservative movement had effectively co-opted his critique of the press for their own purposes~\cite{bauer2026making}.

AIM marks the beginning of the institutionalization of a critique that both rests on and sustains the assertion that the media is liberal~\cite{bauer2026making}. Such research is often funded by big business, which, as a rule, prefers not to be investigated, and is therefore generally hostile to an adversarial press~\cite{Herman1999Myth}. By the time that the Center for Media and Public Affairs (CMPA) began publishing its \textit{Media Monitor} in 1985, this critique was fully institutionalized. CMPA's research aimed to ``demonstrate the liberal bias and anti-business propensities of the mass media''~\cite{bigchomsk}. Their work often employs dubious methodologies designed to reach the predetermined outcome:

\begin{quote}
    [T]he main analytical technique used by the Center—the counting of “thematic messages”—is extremely dubious, eliminating all messages that fail to make an explicit statement of opinion. Since sources who accept the status quo don’t need to explicitly state an opinion, this technique often produces highly distorted findings. For example, the CMPA report on Gulf War coverage found that ``nearly three out of five sources (59 percent) criticized U.S. government policies during the Gulf War.'' This improbable result comes from throwing out 5,666 of 5,915 messages, and looking only at what the remaining 249 said about U.S. policy.~\cite{hartMeetMythMakers1998}
\end{quote}

CMPA, AIM, and other think tanks continue this work to the present day~\cite{hartMeetMythMakers1998, Herman1999Myth}.

In the 1990s, Rupert Murdoch founded his network, Fox News, explicitly to combat the supposed liberal bias of the media, as did other conservative media moguls, like Roger Ailes and Conrad Black, all of whom built influential media empires~\cite{Herman1999Myth}. With Murdoch, we see the concept of the liberal media become fully mainstream. This mainstream acceptance provides credibility to the founding of explicitly rightist media organizations aiming to balance liberal media, even as rightist media organizations increasingly dominate the media industry~\cite{pew2023cablenews}.

As we saw in Section~\ref{sec:data}, MBFC's data reflects this intellectual lineage, consistently rating mainstream media organizations as liberal by looking at its content. Their ratings do not, however, contain a critique of capital, such as those of the early Progressive Era reformers. MBFC's data is downstream of this history, but uncritically so, and presumably entirely unaware of it.

\subsection{A critique of MBFC's methodology}
\label{sec:stop}
As we show in the previous section (Section~\ref{sec:misconceptions}), there are widespread misconceptions about MBFC in the misinformation literature. Similarly, as we discuss in Section~\ref{introduction}, university libraries recommend MBFC, and it is increasingly integrated into consumer products. We therefore consider its methodology here.

In Section~\ref{sec:data}, we provide descriptions of MBFC's data, and explain the rubrics MBFC uses in generating ratings. MBFC's rubrics make clear that its data is the result of extensive interpretive work. This interpretive work, however, does not meet basic academic standards. Our argument here is not that media outlets cannot be quantitatively compared or that measurement is a lossy process. The problem here goes beyond this inherent property of quantification. Put simply, MBFC's methodology is sloppy.

Recall the arbitrariness in point values found in Section~\ref{sec:data}. Each score is composed of sub-scores, which are given percentage weights without explanation, e.g., \mbfcvar{bias} is 35\% \mbfcvar{economic system}. Similarly, scores are generated, then transformed into bins of arbitrary sizes, again with no explanation, and only those secondary, transformed values are made available to the end user. The \mbfcvar{credibility} rubrics contain nested logic without a theoretical justification. The \mbfcvar{bias} values outside the left-right spectrum similarly typify a poorly conceptualized rubric at all levels, from the measurement values themselves to the definitions it uses in its rubrics.

Recall from Section~\ref{sec:data} and Table~\ref{tab:factual-reporting} that the \mbfcvar{factual reporting} score contained within it \mbfcvar{one-sidedness}, the best score for which was a lack of bias. The rubric uses the term ``bias,'' but this use has no explicit relationship to \mbfcvar{bias}, which is a seemingly independent measure, though there is clear conceptual slippage between the pair. This means that MBFC's interpretation of \mbfcvar{factual reporting} contains within it a problematic normative claim that centrist reporting is, by definition, more factual. This conceptual slippage goes both ways, as the \mbfcvar{bias} score contains values that live entirely outside the \mbfcvar{bias} spectrum, but instead seem more like questions of \mbfcvar{factual reporting}.  In short, though the rubrics that MBFC uses often clearly spell out procedures, these procedures are arbitrary.

In the rare cases that we do get theoretical discussions, they are inadequate. Consider again Table~\ref{tab:economic-system}, which contains the rubric for the \mbfcvar{economic system} portion of the \mbfcvar{bias} score. MBFC's methodology page gives the appearance of an academic article: It has descriptions, rubrics, mentions of constraints, and a references section at the bottom. Though it satisfies these aesthetic expectations of rigorous work, it deploys these aesthetics without substance. The page contains 18 references at the bottom without in-line citations to indicate from which source(s) any particular definitions are drawn~\cite{Methodology2026}.  Upon searching through them, we do not find the MBFC definitions in any of the citations, nor anything topically relevant. These references are styled similar to an academic paper's references, but they are in fact more like a ``further reading'' section, and contain no citations supporting the methodology, as one would expect. This lack of citations creates problems of integrity and interpretation for scholarship downstream of MBFC, though it would not be apparent upon skimming the page.

Despite having no meaningful citations, many of the definitions in the \mbfcvar{economic system} rubric are contentious, and some have well-documented historical and current counterexamples. 

Recall from Table~\ref{tab:economic-system} that the furthest left \mbfcvar{economic system} 
``[a]dvocates no corporatism, extreme regulation, and full government ownership of industries.'' 
Nazi Germany practiced an economic doctrine known as \textit{Gleichschaltung} (roughly ``total coordination'') with the state, requiring that ``endless accountings be submitted regularly to government bureaus'' and ``companies install Hollerith machines [an IBM punch-card computer] to ensure prompt, up-to-the-minute reports''~\cite{black2002ibm}. 
Fascist Italy underwent similar processes: ``State intervention in the economy blurred the lines between the private and public sector to such a degree that employers were in fact transformed into such agents of the state''~\cite{elward_fascist_corporativism_2023}. 
Under MBFC's rubric, these economic systems would classify as leftist. Since \mbfcvar{economic system} is 35\% of the total \mbfcvar{bias} score,  it would be mathematically difficult for the final \mbfcvar{bias} score of an outlet that championed Mussolini's Italy or Hitler's Germany to be further to the right than the center. 

For a more current example, roughly 40\% of Saudi Arabia's GDP comes from oil~\cite{forbes_saudi_arabia} extracted by a state-owned company~\cite{saudiaramco_who_we_are_2018}. 
According to the rubric, this too is a leftist system, so Saudi propaganda would, we expect, tilt left. MBFC's ratings for \textit{Arab News}, which, according to MBFC, is propaganda for Saudi Arabia, rates the source \mbfcvarsecond{right-center}.
MBFC provides a blurb describing their reasoning, and \mbfcvar{economic system}, despite supposedly making up 35\% of the score, is not mentioned anywhere on the page~\cite{mbfcarabnews}. 

Similarly, MBFC's data rates the \textit{Huffington Post} as far left as the scale allows. It is difficult to imagine that \textit{Huffington Post} advocates anything like ``no corporatism, extreme regulation, and full government ownership of industries,'' and MBFC's write-up again makes no mention of \textit{Huffington Post} as a mouthpiece for the international proletariat, instead focusing on the outlets attacks on Donald Trump~\cite{mbfc_huffpost}. 

These examples are reasons to doubt the consistency of application of MBFC's rubrics, but recall also from Section~\ref{sec:data} that we are quoting the new rubric (from 2025). If the rubrics are objectively applied, results from before and after 2025 should not be comparable. Unfortunately, there is no archived version of the pre-2025 data, so quantitative comparisons are impossible.

There is, however, no satisfactory outcome: If the rubrics are objectively and procedurally applied, then studies from before and after are not comparable. If they are not consistently applied, then that is a problem on its face. We have not found a paper in our database that mentions this 2025 rubric change, and the majority of the papers in our database are from before the rubric update, meaning that they were relying on MBFC data before it had even the stylings of rigor.

Setting aside these inconsistencies, as well as historical and extant counterexamples, MBFC's definitions also invoke yet contradict theoretical traditions without explanation. Communists and socialists have defined these terms for centuries~\cite{engels1847principles}. Rather than a society in which the government owns everything, as MBFC describes it, communism is, according to communists, a stateless society, one that would arrive at the end of a transition period of worker control over the means of production~\cite{marx1875gotha,lenin1917state}. Communist parties in power have maintained this distinction, viewing their regimes as building towards communism~\cite{nguyenphutrong2021socialism}, or in the ``primary stage''~\cite{prcconstitution2023}. 

The right of the spectrum, meanwhile, seems to contain the ideas of the free market from ~\citet{friedman1962capitalism} or ~\citet{hayek1976freemarket} (though, without citations, we cannot be sure). Unlike the dismissive posture taken towards the aforementioned socialist tradition, free market absolutism is presented on its own terms. The rubric does not engage with arguments that markets are not spontaneous, but political creations~\cite{polanyi1944great, thompson1963making}, often coerced into being with military force, and, in the 20$^{th}$ century, often through US intervention, in opposition to democratically-elected governments~\cite{bevins2020jakarta,klein2007shock,graeber2011debt,prashad2020washington}.

This description of the right also diminishes the variety of its political traditions. Charles de Gaulle's government, for example, undertook significant ``indicative planning'' (i.e., state intervention) in the economy~\cite{golub2025degauleeconomicpolicy}, and did not ``[govern] entirely by free-market principles and private enterprise.'' Even in the United States, the rise in military expenses by right-wing governments~\cite{coletta2021defensebudget, houseappropriations2026pentagonbudget} exemplifies a more complex relation to state intervention in the economy than this rubric's simplistic definition~\cite{cypher2015militarykeynesianism}.

Recall also, from Section~\ref{sec:history}, as part of a rightist attack on the media, CMPA deployed a dubious methodology to attempt to propagate the liberal media claim. MBFC's methodology contains echoes of the same. In the methodology page's introduction, MBFC writes:

\begin{quote}
    Special attention is given to detecting bias by omission, one-sided narratives, and the use of unreliable sources~\cite{mbfc}.
\end{quote}

We saw this attention in Section~\ref{sec:bias}'s \mbfcvar{straight news reporting balance}, which...

\begin{quote}
     [m]easures how well a source reports all sides in its straight news stories, either through story selection or content balance within articles. This covers strictly news reporting and is separate from Editorial/Op-Ed bias.
\end{quote}

Though not identical, ``content balance'' echoes CMPA's ``thematic messaging'' and other, similar techniques used by various entities discussed in Section~\ref{sec:history}. We suspect that this is an example of how MBFC has been influenced by much of the history described in Section~\ref{sec:history} but, again, without citations, we cannot know.

The point here is not asymmetry, hypocrisy, or incompleteness, but lack of rigor. MBFC presents rubrics absent theory or meaningful, specific citations to the literature. If scholarship is a cumulative endeavor that grows and branches, then scholarship using MBFC develops and belongs on a disconnected branch. MBFC's own site confirms this observation, as we saw in the disclaimer quoted in Section~\ref{sec:misconceptions}.

In the next section, we theorize why, despite methodological flaws, and even a disclaimer, MBFC can be found in studies of media and (mis)information throughout the literature.

\subsection{Why studies use MBFC}

In previous work~\cite{ruiz}, we found an explosion of the term ``misinformation'' in academic literature beginning in 2016, when ``Trump [pulled] off biggest upset in U.S. history''~\cite{politico2016electionresults}. This happened as Americans increasingly used social media~\cite{pew2025socialmedia}, thus creating an enticing dataset of political discourse through which to study this recent and unexpected phenomenon of great public interest. 

As one study writes, Trump's opponent had promised to ``follow the science,'' whereas the Trump administration planned ``to undermine science,'' making this a phenomenon of particular interest among the scientific community~\cite{webb2022politicsscience}. Scientists, presumably aware of the political stakes of appearing biased themselves, turned to data for its presumed objectivity~\cite{porter}, made available through the aforementioned adoption of social media, and settled on a framing that centered facts, that of misinformation~\cite{ruiz}.

This sets up a tension between the vast, computer-legible corpus of social media posts and the innate ambiguity involved in extracting concepts like truth or bias from snippets of natural language, especially at scale. MBFC allows research to resolve this tension.

MBFC fills the gap between the questions that data analysis can answer and the plainly subjective objects of inquiry by creating a corpus of these plaintly subjective variables, allowing research to simply use MBFC's data without having to insert their own subjectivity. 

When doing their exploratory analysis of MBFC, researchers find data that, presumably, matches what they expect, e.g., that the media is center-left and mostly reliable (Section~\ref{sec:bias}), or that official enemies of the US are less free (Section~\ref{Freedom Index}). They might take this agreement as proof that MBFC is reliable, when instead it is proof that they are both downstream of the same political processes. MBFC's success in the literature, we argue, comes not from its accuracy, but from its faithful quantification of hegemony. MBFC makes the dominant ideology legible to computational methods. It allows research to superficially remove subjectivity by accepting hegemony.

\citet{linHighLevelCorrespondence2023} find that MBFC has a high correspondence with other, similar data sources, such as NewsGuard and Ad Fontes. This is not a coincidence. Each of these different sources, in some form or another, aim for neutrality by sampling average people. In the case of MBFC, $N=1$ (though MBFC does have volunteers), but Ad Fontes, for example, asks a panel of one conservative, one liberal, and one centrist to rate sources after attending a training~\cite{adfontes_methodology}. As we saw in Section~\ref{sec:history}, in politics, ``common sense'' ideas are downstream of successful political projects, meaning that methods that attempt to be neutral by gauging the general public find hegemony, not neutrality~\cite{gramsci1971prison}. 

This problem runs deeper than the final, downstream concepts. We now know the history of the liberal media, but the media is also, in the philosophical sense,  a quintessentially liberal institution~\cite{greenbergIdeaLiberalMedia2008b}. It is sometimes called the ``Fourth Estate,'' a reference to the Estates-General and the French revolution, perhaps the paradigmatic liberal revolution~\cite{tocqueville1856ancien}. The ``liberal'' in liberal media, however, refers to a uniquely American usage of the word, in which there is a center, and ``liberal'' means left of center. In Australia, by contrast, the Liberal Party is referred to as a conservative party in headlines without controversy~\cite{chen2025australias}. 

Similar to the battle to label the media as liberal, the meaning of words like ``liberal'' are active terrains of struggle, as are all words used in politics, because power is wielded through language~\cite{gramsci1971prison, hume1777firstprinciples}. For scholarship seeking simple, neutral data on politics~\cite{porter2}, this is not a resolvable problem, because, in politics, the definition of every word will be forever contested. New concepts will get mapped onto old words while old concepts get warped, twisted, and discarded as different coalitions jockey for rhetorical position. 

As we saw in Table~\ref{tablepapers}, a quarter of the most cited papers leveraging MBFC are attempting to do misinformation detection, or to evaluate the veracity of statements at scale on social media through natural language processing of text. If the stated goal of misinformation research is to intervene in some kind of political malady, as many papers state explicitly~\cite{ruiz}, then they face a serious challenge, because political struggle happens by and through words. In the process, words and meanings change, and they risk being trapped and confused by these rhetorical maneuvers.  

MBFC allows researchers to study politics, if superficially, while actually side stepping it. At first glance, MBFC's data mirrors a common practice in machine learning, that of human or expert annotation of ``ground truth'' to train models. Commonly used datasets like MNSIT, for example, provide photographs of hand-written integers~\cite{lecun1998gradient}, and a typical workflow might include human annotation for each image, then training a model to classify new, unannotated photos of integers.

In our case, instead of dealing with inexact language, researchers have a dataset that has, through pseudo-scientific annotation, extracted politics from its native medium. The resulting data would allow them to view politics clearly, using numbers, as if from the outside, rather than bogged down in rhetorical mud. The error is that the inexactness of language is where much political work happens. Outside political struggles, where the definitions of words are less contested, such annotations are on firm ground. When using MBFC, however, in an attempt to climb out of politics, researchers entrench themselves further, inadvertently accepting hegemony.  

Our case is an extreme example of a generally unavoidable fact of data. The natural world does not come with a quantified representation of itself. Data comes from measurements, and someone with their own ideas, interests, and flaws has to take those measurements. This is, in and of itself, neither good nor bad. It is just true, and means that all quantitative results are necessarily derived from and embedded in a qualitative understanding of the world. This paper is a case study in what happens when research does not acknowledge this fact. We explore the consequences of this error in our next and final section, where we implore researchers to stop using MBFC.

\section{Conclusion}
\label{sec:conclusion}

\begin{figure}[t]
    \begin{center}
    \includegraphics[width=\linewidth]{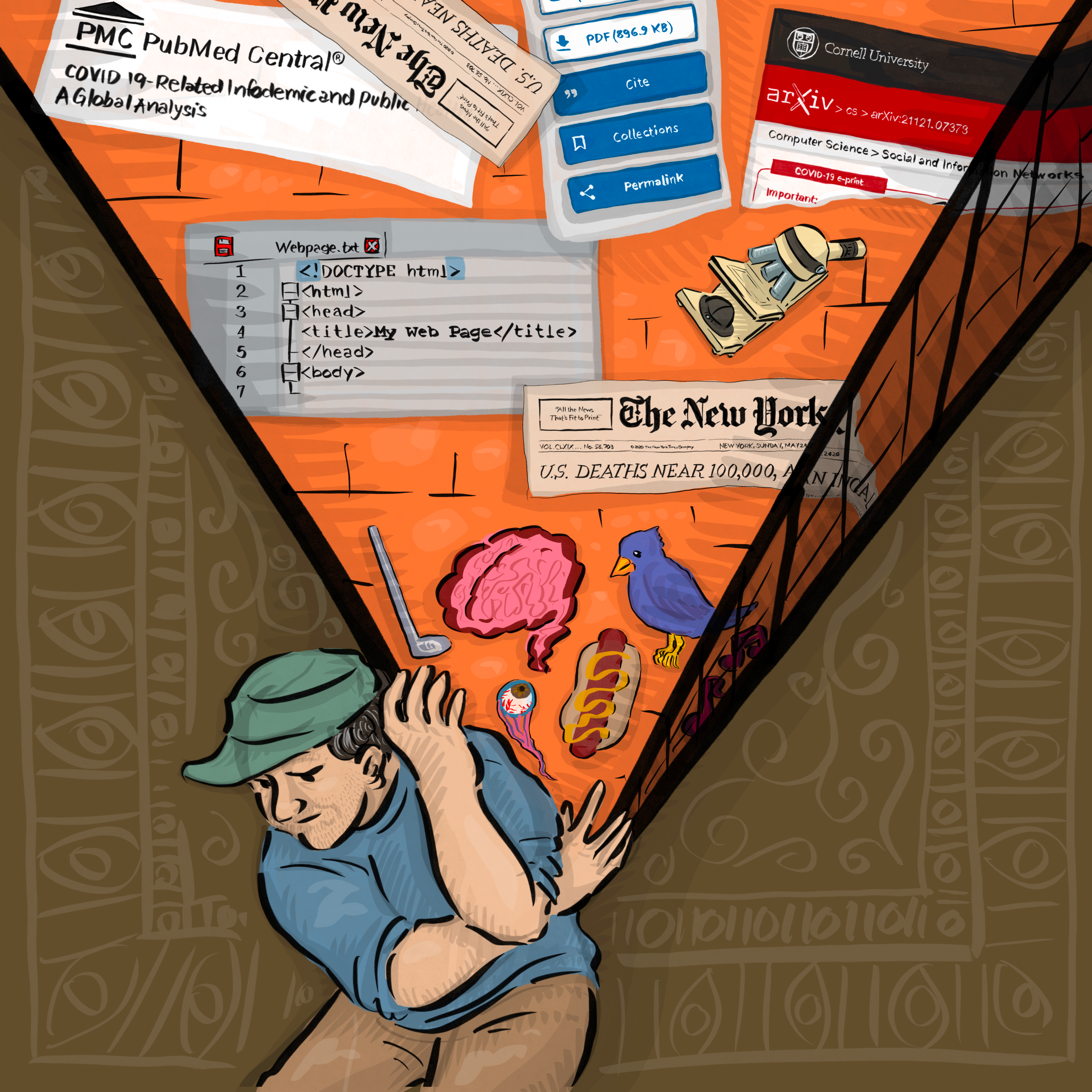}
    \caption{An illustration of the single point of reliance on MBFC for ground truth on the credibility and bias of news outlets---a ground truth used in various research contexts and commercial products.}
    \label{fig:washingpyramid}
    \end{center}
  \end{figure}
  
In this study, we have shown that a substantial amount of published misinformation scholarship relies on MBFC. Despite often using MBFC to rate the bias and reliability of sources, misinformation scholarship consistently misrepresents MBFC itself, one of its most used sources. The acute public interest in the topic, the available dataset of social media, the need to bridge the limitations of that data, and the scientific desire for objectivity and neutrality created the conditions for MBFC's popularity. 

The core of problem, however, exists outside MBFC. This study is highly critical of MBFC, but this criticism is aimed at the scholarship that relies on it. Good scholarship need not come from the academy, but academic institutions exist to preserve and expand good scholarly practice. Researchers who wanted to use MBFC might have offered productive feedback, or even collaboration. Instead, the prevalence and use of MBFC in peer-reviewed literature during waning public trust in science has become an ironic case study.

As we saw in Section~\ref{sec:stop}, MBFC takes for granted the results of a decades-long conservative attack against the media, a propaganda campaign to which the entire Anglosphere has been subjected. MBFC then quantifies this rhetorical victory as data and, aided by the work of misinformation scholars, it is inserted back into discourse as scientific fact. 

In a sense, it is true that the media is liberal, but only because it was made true through a conscious political project to discredit it. Now, throughout the misinformation literature, in its attempt to understand and alleviate: 
``[t]he explosive growth in fake news and its erosion to democracy, justice, and public trust''~\cite{Alonso2021Sentiment}; 
``distrust in scientific expertise''~\cite{Lenti2023Global}; 
``political partisanship and mistrust of science''~\cite{Hu2021Socioeconomic}; and
``distrust in science [that] undermines public health and may drive civil unrest''~\cite{Broniatowski2023Facebook}, 
misinformation scholarship has ironically turned this successful attack into a scientific consensus (each of these research motivation quotes comes from a paper using MBFC). This consensus then buttresses the putative \textit{raison d'être} of conservative media like Fox News, whose ``Fair and Balanced'' slogan refers not to its own objectivity, but to its conservatism balancing the liberal media~\cite{bauer2026making}.

As we also saw, many of the papers in our database seek to detect misinformation on social media to alleviate a crisis of credibility or institutional trust. In the process of doing so, they unwittingly encode the arguments through which the rightist media ecosystem justifies itself into technical systems that moderate public discourse. 
In this, misinformation scholars make a mistake with historical precedence. When Morris Ernst wrote his structural press critiques,  

\begin{quote}
...the common sense among postwar media reformers was that the concentration of media ownership would disproportionately stifle liberal voices [...] [W]hen Ernst agreed to join the AIM board [...] the news media now was increasingly considered to be controlled by a cabal of liberals~\cite{bauer2026making}.
\end{quote}

Ernst's sincere if na\"ive commitment to truth in media allowed him to be used by political actors to legitimize themselves. Many journalists made the same mistake:

\begin{quote}
    AIM's thin veneer of impartiality---achieved in part through the group's affiliation with Morris Ernst---provided enough plausible deniability that some corner of the journalism profession felt compelled to take it seriously~\cite{bauer2026making}.
\end{quote}

Misinformation scholarship's search for a fair, neutral procedure by which to moderate public discourse leaves it vulnerable to co-option by unscrupulous and well-funded political actors, some of whose propaganda campaign it already renders as scientific fact. Just as there is no definition of ``liberal'' that does not take a political stance, there can be no neutral dataset or detection scheme. So long as truth has partisan opponents, to be for truth is a partisan stance. Ernst's error, then, should be a lesson in the dangers of na\"{i}vet\'{e} in the face of politics. It cannot be the role of scholarship to feed the output of a political process back into itself, detached from context and laundered through academic prestige. We therefore urge researchers to stop using MBFC.

\bigskip
\section{Acknowledgments}

Earlier drafts of this paper predated the publication of AJ Bauer's wonderful \textit{Making the Liberal Media}, and, without its aid, the historical discussions of those drafts left much to be desired. We therefore express our gratitude for his work, and congratulate him on the completion of an excellent, thorough, and necessary book.

We thank Simon Tremblay-Pepin for his invaluable feedback on the first draft. Thanks also to Tabia Tanzin Prama and Calla Beauregard for their enthusiastic support of this project.

The authors acknowledge support by 
National Science Foundation awards \#2419830 and \#2242829 
and MassMutual.

\bibliography{refs}

\clearpage

\end{document}